\documentclass{elsart}

\usepackage{graphicx}
\usepackage[T1]{fontenc}

\usepackage{amssymb}

\begin{document}

\begin{frontmatter}

\title{Fraudulent agents in an artificial financial market}

\author[DISTA]{Enrico Scalas\thanksref{email1}}
\thanks[email1]{Corresponding author. E-mail: scalas@unipmn.it}
\author[DIBE]{Silvano Cincotti}
\author[DIBE]{Christian Dose}
\author[DIBE]{Marco Raberto}

\address[DISTA]{DISTA, Università del Piemonte Orientale, Corso Borsalino 54, 15100 Alessandria,
 Italy}
\address[DIBE]{DIBE, Università di Genova, Via Opera Pia 11a, 16145 Genova, Italy}

\begin{abstract}
The problem of insider trading and other illegal practices in
financial markets is an important issue in the field of financial
regulatory policies. Market control bodies, such as the US SEC or
the Italian CONSOB \cite{Minenna01}, regularly perform statistical
analyses on security prices in order to unveil clues of fraudulent
behaviour within the market. Fraudulent behaviour is connected to
the more general problem of information asymmetries, which had
already been addressed in the field of experimental economics
(see, for instance, refs. \cite{Plott82,Forsythe90,Huber03}).
Recently, interesting conclusions were drawn thanks to a
computer-simulated market where agents had different pieces of
information about the future dividend cash flow of exchanged
securities \cite{Chan01}. Here, by means of an agent-based
artificial market: the Genoa Artificial Stock Market (GASM)
\cite{Cincotti03,Marchesi03}, the more specific problem of
fraudulent behaviour in a financial market is studied. A
simplified model of fraudulent behaviour is implemented and the
action of fraudulent agents on the statistical properties of
simulated prices and the agent wealth distribution is
investigated.

\end{abstract}
\end{frontmatter}

\section{Introduction}

Laws and regulations define fraudulent behaviour in a financial
market. Among the many possible instances of fraud, we will focus
our attention on insider trading \cite{Minenna01,Bhattacharya02}.
It must be stressed that the enforcement of insider trading laws
took place mainly in the 1990s, as a consequence of advances in
the economic theory as well as in the empirical analysis of
financial markets. Actually, many papers have been devoted to
insider trading: a recent review by Bainbridge, quoted by
Bhattacharia and Daouk, lists 261 papers \cite{Bhattacharya02}. In
the past, various arguments were presented against the
introduction and enforcement of insider trading laws. For
instance, it has been pointed out that exploiting insider
information is the only way to compensate managers. It has been
also argued that the insider pushes the stock price faster towards
a value, which better reflects the fundamental value of the
company. Although the problem of evaluating the social and
economical effects of insider trading can only be assessed by
means of accurate empirical studies, there are several
difficulties to be overcome, mainly due to the fact that several
pieces of information on insider trading are confidential.
However, after an empirical analysis of stock markets in 103
countries, Bhattacharya and Daouk were able to conclude that
enforcement of insider trading laws reduces the cost of equity by
about 5\% \cite{Bhattacharya02}.

In European countries where insider-trading laws are enforced, the
scope of the law includes any security listed on a stock exchange,
but also face-to-face transactions may be taken into account. The
inside information is defined as non-public and price sensitive.
Among the enforced subjects there are institutional insiders,
other primary insiders and tippees. Some forbidden behaviours are:
trading on own account, trading on behalf of third parties and
communicating the information to third parties. There may be
penal, civil and administrative sanctions, including imprisonment
and fines. Fines are computed in relation to the disgorgement: the
economic value of the insider information.

There are several methods to compute the disgorgement. The US
Security Exchange Commission uses a procedure, which is based on
the Event Study Analysis in order to evaluate the return
percentage variation caused by the insider information. This is
called potential econometric disgorgement. To avoid some pitfalls
of the US method, the Italian body (CONSOB) has developed a new
probabilistic approach, which will be discussed below.

Insider trading is an instance of information asymmetry in a
financial market, which was discussed at length in the literature
\cite{Plott82,Forsythe90,Akerlof70,Fama70,Grossman80,Fama91}.
Recently, it has been addressed by means of an artificial market
[4]. With artificial markets, it is possible to study the effect
of information asymmetries in a well-defined and controlled
environment.

In this paper, we present a simplified model of insider trading in
the Genoa Artificial Stock Market (GASM). The paper is organised
as follows. Section 2 is devoted to the methodology of
probabilistic disgorgement and to clarify the procedure used here
to compute disgorgement in the artificial market. In section 3,
the results of the GASM simulations are presented and discussed.
Finally, in section 4, the reader can find the main conclusions
and the direction for future work on insider trading in artificial
financial markets.

\section{An alternative path to probabilistic disgorgement}
As mentioned in the Introduction, the Italian market regulation
authority (CONSOB) has recently developed a new approach to unveil
the economic value of the information exploited by each insider.
CONSOB has called this procedure potential probabilistic
disgorgement \cite{Minenna01}. The (simplified) procedure consists
of the following steps.

At first, two time-horizons are defined, $\alpha$  and $\Theta$.
The former corresponds to the period in which the insider will
build his/her position on the stock; the latter is defined by the
day in which the insider information is published and the first,
the second or the n-th day after. The number of days considered is
related to the liquidity of the stock under investigation and, in
a standard insider-trading scheme, it coincides with the instant
in which the insider closes his/her position.

The second step rests on two hypotheses. The first one is that the
insider cannot control the price dynamics before the occurrence of
the event and the second one is that he/she will create a long
(short) position, if the event information will induce an increase
(decrease) of the stock price. In other words, the insider
predicts that the preferential information will move the market
more in the period $\Theta$  than it moved in the period $\alpha$.
Therefore, the parameters of a geometric Brownian motion
representing the time evolution of the stock in the period
$\alpha$ are estimated, by using daily data.

Finally, on this basis, an oscillation band for the price of the
stock under study in the time-period   is defined by fixing a
significance level. The difference between the actual stock price
after the insider information is revealed and this band represents
the disgorgement.

A slight but important modification in computing the potential
probabilistic disgorgement would be replacing the geometric
Brownian motion with a process that better represents the
leptokurtic behaviour of the return probability density estimate.
To this purpose, there are various possible choices. An
interesting possibility would be using truncated Lévy walks {\it à
la} Koponen \cite{Koponen95} or their generalisation proposed by
Boyarchenko and Levendorskii \cite{Boyarchenko00}. Here, however,
as we use an artificial market, we can directly simulate the
behaviour of the same market, with and without the action of the
insider. Therefore, in the next section, we shall directly
evaluate the impact of insider-trading activities.

\section{GASM model for insider trading}

\subsection{Brief description of the artificial market}

We modelled the behavior of an insider trader in the framework of
the Genoa Artificial Stock Market (GASM).  In the market,
$N=10,000$ random traders buy and sell a risky asset in exchange
of cash. The decision-making process of each random trader is
constrained by the limited financial resources and influenced by
the volatility of the market. The resulting price process is
characterized by reversion to the mean and exhibits fat-tailed
distributions of returns and volatility clustering
\cite{Raberto01}. The average price (equilibrium price) is given
by the ratio of the total amount of cash present in the system and
the number of shares of the traded stock. In a closed market, i.e.
no cash or shares inflow or outflow in the system, the long-run
price average is constant. An increase of the global amount of
cash in the system causes an equal increase of the long-run mean
value.

\subsection{A model of insider trading}
\label{Sec:model} The price process exhibits a mean-reverting
behavior around the equilibrium price; thus a fundamentalist
trading strategy with fundamental price equal to the equilibrium
price would be profitable \cite{Raberto03}. As price sensitive
event, the injection of liquidity in the market has been chosen,
simulating the action of a central bank. Being the equilibrium
price equal to the ratio between the total amount of cash and
share number, liquidity injection yields a new higher equilibrium
price. A fundamentalist agent, who knows in advance the date of
this event, should be able to take profit of this piece of insider
information.

We consider a number of different simulations with the following
features:
\begin{itemize}
    \item $N$ random traders and 1 insider fundamentalist trader operate in the
    market; at each time step, each random trader issues orders
    with probability $q$;
    \item at the beginning of the simulation ($t=0$), each random trader receives the same
    amount of liquidity $c(0)$ and number of shares $a(0)$, while the insider
    is given liquidity $\hat{c}(0)$ and zero number of shares $\hat{a}(0)$;
    \item the initial value of the stock price is set at the
    equilibrium price:
    $$\bar{p}=(Nc(0)+\hat{c}(0))/Na(0) \eqno{(1)}$$
    in order to quickly reach the equilibrium
    state characterized by fluctuations around the equilibrium price;
    \item background trading is realized by the action of random
    traders, who issue random buy and sell orders with no
    information on the timing and magnitude of liquidity injection;
    \item at a given time $t^*$ the global amount of cash is
    increased by 10\% and the new cash is distributed to agents
    proportionally to the fraction of the total wealth
    (cash+number of shares) they own. Thus the equilibrium
    price increases by 10\%:
    $$\bar{p}(t\ge t^*)=(1.0+0.1) \cdot \bar{p}.$$
    Indeed, for $t < t^*$, the equilibrium price is given by eq.
    (1);
    \item the insider knows the timing and magnitude of
    the injection of liquidity in advance; let $\tau$ denote this advance;
    \item the insider is inactive for $t<t^*-\tau$ and acts
    as a fundamentalist trader with the new appropriate fundamental
    value, $\bar{p}(t\ge t^*)$, for $t^*-\tau \le t < t^*$. This means
    that in this time period, the insider tries to convert all the
    cash into stocks. For $t\ge t^*$, the insider issues sell
    orders if the stock price is greater than the fundamental
    value $\bar{p}(t \ge t^*)$, otherwise he/she keeps its position.
\end{itemize}

As a final remark, the period $\alpha$ discussed in Section 2
corresponds to $t^* - \tau \le t < t^*$ when the insider builds
his/her position, whereas, the periods $\Theta$ in which he/she
leaves the market corresponds to $t \ge t^*$.

\section{Simulations}

As already mentioned, the number of random traders operating in
the simulations is 10,000. The probability of activation is
$q=0.02$. The initial amount of liquidity is $c(0) = 100,000$
units, whereas the initial share number of the risky stock is
$a(0) = 1,000$. As for the insider, his/her activity is
characterized by two parameters: initial liquidity, $\hat{c}(0)$,
and advance, $\tau$. The chosen values for $\hat{c}(0)$ are:
$$\hat{c}(0) = 2^k \cdot c(0), \; k = -1, 0 , 1, 2, 3, 4. $$
The chosen values for the advance $\tau$ are: 20, 40, 60, 80, 100
time steps. For each value of $\tau$ and $\hat{c}(0)$, 25
independent realizations have been simulated, for a total of 750
simulations. Each simulation lasts 2,000 time steps, which in the
framework of GASM can be considered as trading days. The insider
event occurs at $t^* = 200$ days.

In Figure 1, results of a typical simulation are shown. In this
particular case, $\hat{c}(0) = c(0)$ and $\tau = 20$ days. The
left-upper plot displays the price variation as a function of
time. The effect of  the liquidity injection at time $t^* = 200$
days is evident: after the injection, the average price increases
towards 110 units and then fluctuates around this value. The
right-upper plot shows the insider cash as a function of time.
Essentially, the insider starts with 100,000 units of cash and
ends with nearly 120,000 units, with a return of about 20 \%. In a
few days after day $t^* - \tau = 180$, the insider converts all
his/her cash into shares. At day $t^* = 200$, he/she receives the
additional amount of cash like all the random traders. Finally,
when the price becomes greater than the new fundamental value (110
units), he/she starts issuing sell orders and succeeds in getting
rid of the shares in a few days. The behavior of the insider can
be appreciated also in the two bottom plots, where his/her number
of shares (left) and his/her total wealth (right) are presented.

In order to compare the gain of the insider with the average
performance of random traders, Figures 2, 3 and 4 present the
random-trader percentage increase in aggregate cash,
capitalization and wealth, respectively, as a function of insider
cash and advance. The percentage increase is computed between day
2,000 (the end of the simulation) and day $t^* - \tau - 1$ (the
day before the insider entering into the market). The error-bar
half-width corresponds to two standard errors.

In Figure 2, random-trader aggregate cash return is compared with
the gain of the insider. In Figure 3, the insider gain is compared
with the aggregate wealth return of random traders. No definite
trend appears as a function of insider advance, but the insider
average gain systematically decreases with increasing initial
cash. The reason is that the insider influences more and more the
stock price when his/her initial cash grows. In fact, he/she tries
to convert his/her cash into shares as soon as possible, and
issues buy orders that are larger the richer he/she is. Large buy
orders can imply a price increase. If the stock price is higher,
the number of shares he/she can buy for a given amount of cash is
lower. In other words, if he/she influences the stock price,
letting it increase, it becomes no longer true that with a double
amount of cash, he/she is able to buy a double amount of shares.
The decrease in average cash return of random traders is due to
the mechanism of cash distribution within the insider event. When
the wealth of the insider is large enough, random traders receive
less new cash in aggregate (see Sec. \ref{Sec:model}). In Figure
4, the market capitalization return is shown between day 2,000 and
day $t^* - \tau - 1$. No systematic effect due to insider activity
appears. For this reason, also the aggregate wealth return of
random traders in Figure 3 does not show any systematic dependence
on insider parameters.

\section{Discussion and conclusion}
In this paper, a simple model of insider trading in an artificial
market has been presented. Quasi zero-intelligence random traders
are acting on a closed market providing a sort of thermal bath
implying price fluctuations. A fundamentalist insider gets the
information of cash inflow occurring on a future day. He/she tries
to exploit this piece of information by using a fundamental
strategy. It turns out that the insider is able to earn
significantly more money than random traders. Due to the simple
strategy implemented, the excess gain of the insider decreases
when his/her initial endowment of cash is larger. The simulations
presented above show that the Genoa Artificial Stock Market (GASM)
can be used to study instances of fraudulent behaviour in
financial markets. Future work could focus on the effect of
insiders in a market with heterogeneous agents as well as
different price formation mechanisms (clearing house vs book,
...).

\bibliographystyle{IEEEtran}
\bibliography{ecph}

\newpage

\begin{figure}
  \includegraphics[width=1\textwidth]{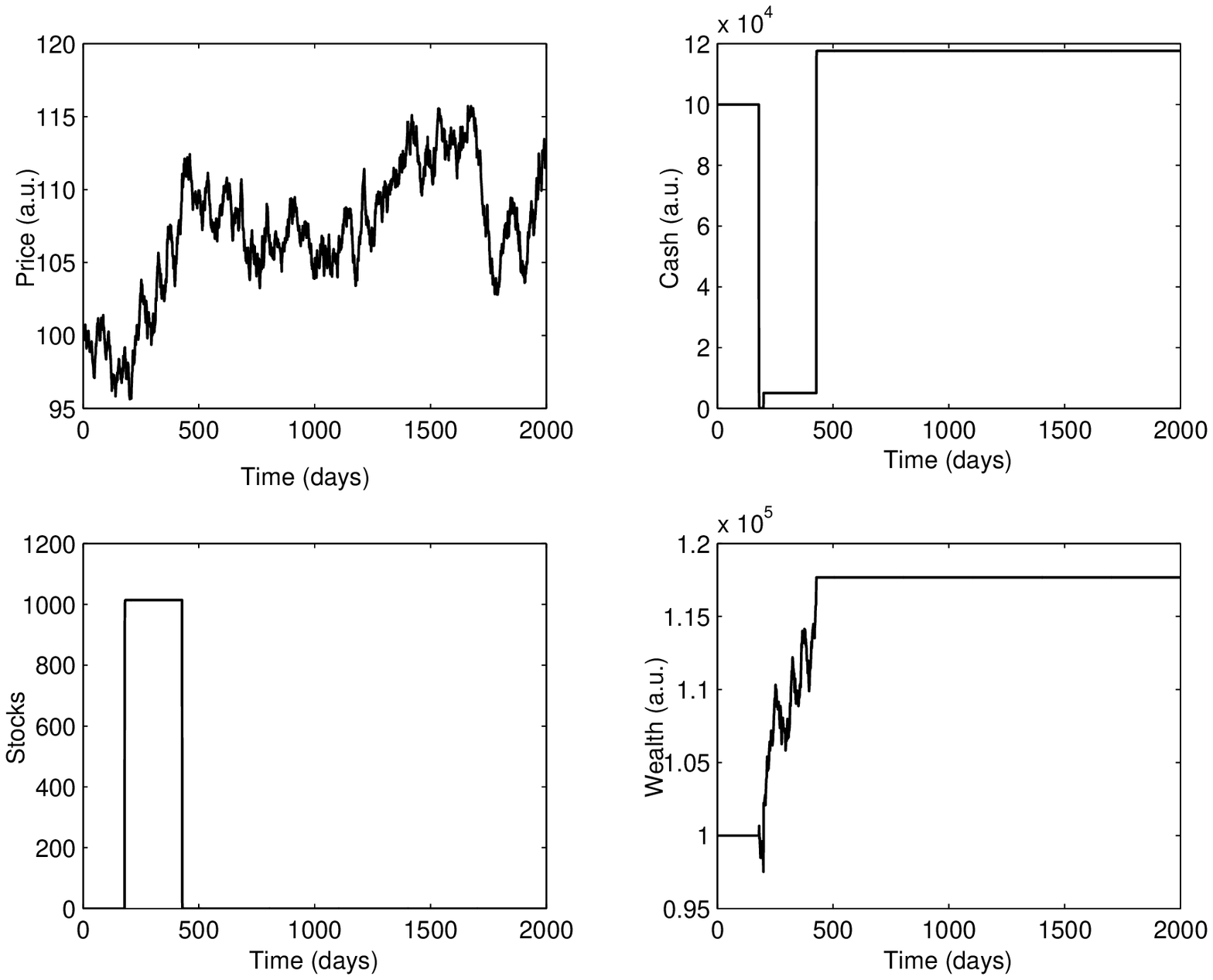}\\
  \caption{Results of a typical simulation.}\label{Fig:1}
\end{figure}

\newpage

\begin{figure}
  \includegraphics[width=1\textwidth]{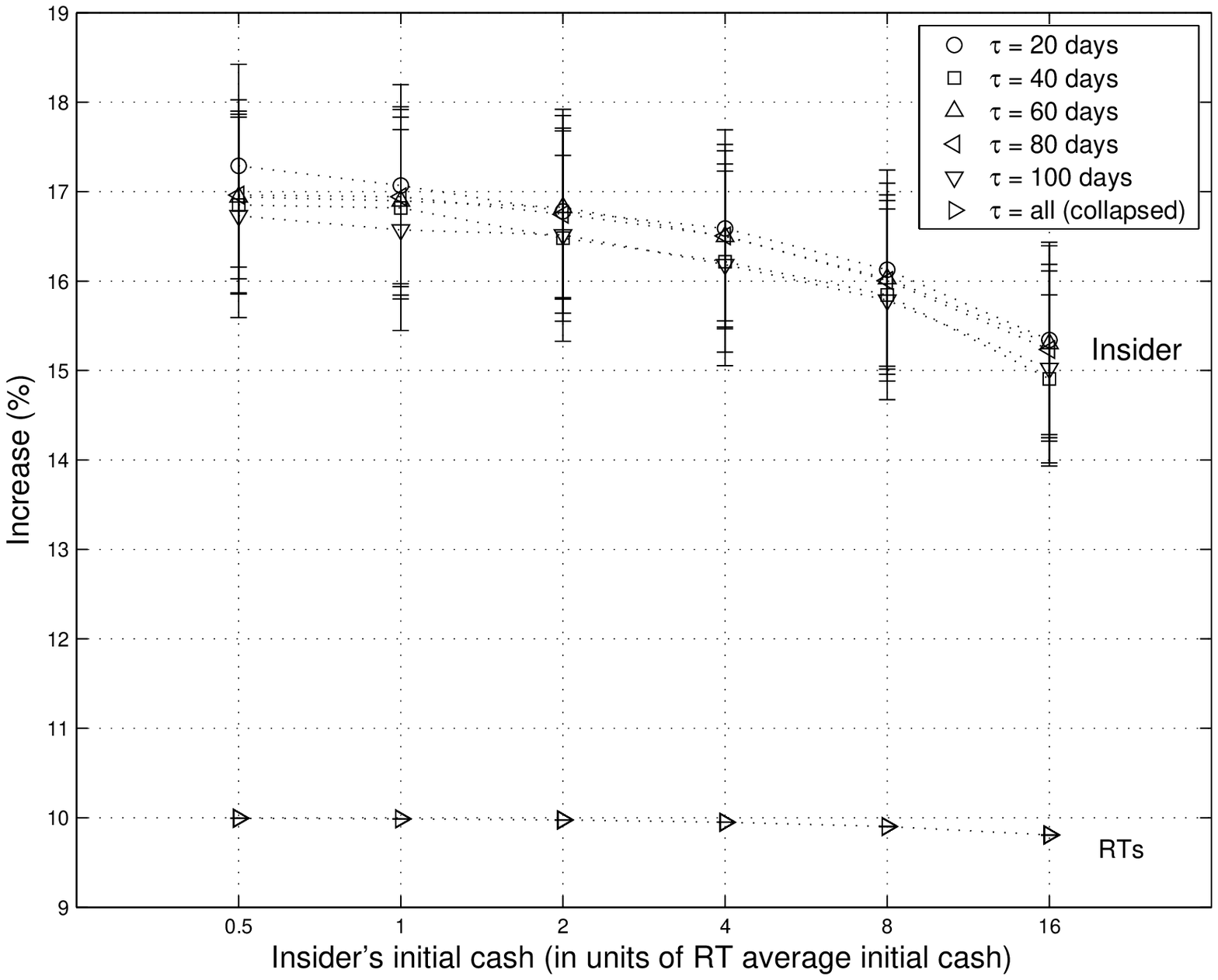}\\
  \caption{Random traders aggregate cash increase compared to insider gain.}\label{Fig:2}
\end{figure}

\newpage

\begin{figure}
  \includegraphics[width=1\textwidth]{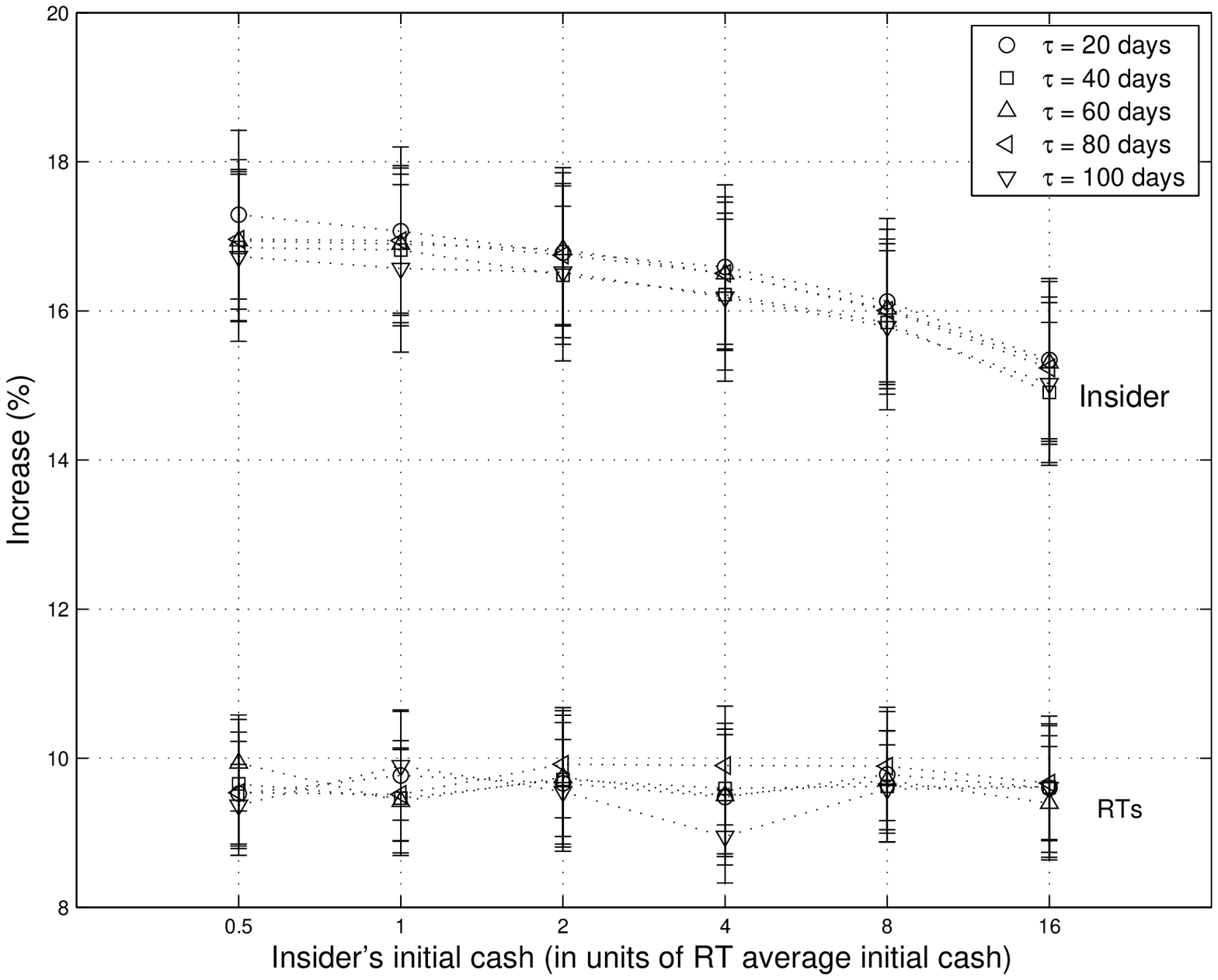}\\
  \caption{Random traders aggregate wealth increase compared to insider gain.}\label{Fig:3}
\end{figure}

\newpage

\begin{figure}
  \includegraphics[width=1\textwidth]{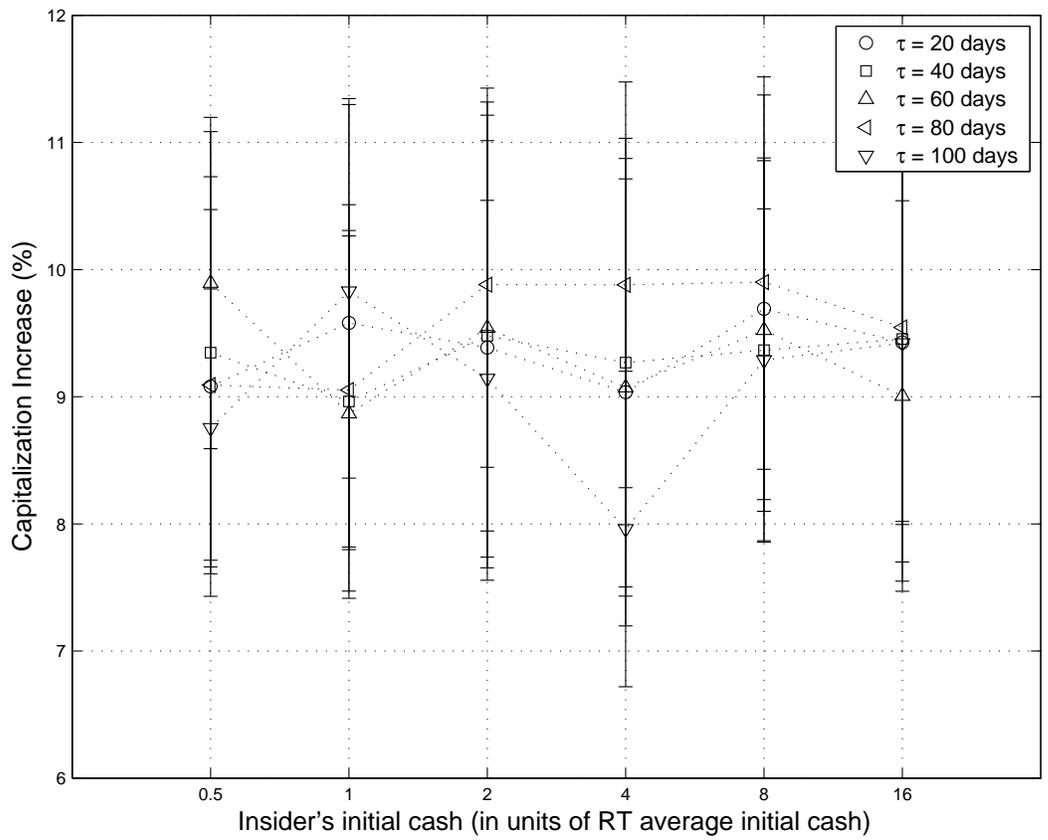}\\
  \caption{Market capitalization percentage increase.}\label{Fig:4}
\end{figure}

\end{document}